# FRAMEWORK TO INTEGRATE BUSINESS INTELLIGENCE AND KNOWLEDGE MANAGEMENT IN BANKING INDUSTRY

G. Koteswara Rao, Indian Institute of Management, Indore, INDIA (gkrao@iimidr.ac.in)
Roshan Kumar, Indian Institute of Management, Indore, INDIA (roshank@iimidr.ac.in)

**ABSTRACT**

In this digital age organizations depend upon the technologies to provide customer-centric solutions by understanding well about their customers' behaviour and continuously improving business process of the organization. Business intelligence (BI) applications will play a vital role at this stage by discovering the knowledge hidden in internal as well as external sources. On the other hand, Knowledge Management (KM) will enhance the organisations performance by providing collaborative tools to learn, create and share the knowledge among the employees. The main intention of the BI is to enhance the employees' knowledge with information that allows them to make decisions to achieve its organisational strategies. However only twenty percent of data exist in structured form, majority of banks knowledge is in unstructured or minds of its employees. Organizations are needed to integrate KM with Knowledge which is discovered from data and information. The purpose of this paper is to discuss the need of business insiders in the process of knowledge discovery and distribution, to make BI more relevant to business of the bank. We have also discussed about the BI/KM applications in banking industry and provided a framework to integrate BI and KM in banking industry.

## INTRODUCTION

Banks worldwide use data warehousing/BI solutions for performance measurement, profitability analysis, risk management, historical analysis, managing compliance requirements, executive dashboards, regulatory reporting and customer relationship management.Once the business transactional data (deposits and loans etc) from all the branches of the banks had been accumulated in the transactional processing system (i.e. a common database in a central server located in the data center), giving a consolidated view of the bank's operations, the need of bank managers at various positions to know the financial status (financial statements, cash flows, and summary reports etc.) of the bank in order to acquire new customers and retain existing customers, was perceived. Therefore, bank needed to develop a platform that would integrate data from various sources within the bank into an easy-to-use, easy-to-locate data delivery service. As a result BI came into existence. BI system brought the perception of knowledge discovery which bankers have quickly adopted for active support to decision making processes at all managerial levels. BI tools founded on information technologies such as on-line analytical processing and data mining make possible intelligent business decision making in complex banking environment. In this paper we have discussed about BI, KM, BI & KM applications for banks, literature review on BI and KM integration, need to integrate BI & KM and a framework for banks to integrate BI and KM.

## APPLICATIONS OF BUSINESS INTELLIGENCE IN BANKING INDUSTRY

A BI system has evident importance as a communication and information diffusion channel, preferably one that is open, trustworthy, transparent and permanent. In supporting the monitoring and evaluation of business results while maintaining information integrity (Petrini and Pozzebon 2009). Zeng et al. (2009) define BI as "The process of collection, treatment and diffusion of information that has an objective, the reduction of uncertainty in the making of all strategic decisions." Experts describe Business intelligence as a "business management term used to describe applications and technologies which are used to gather, provide access to analyze data and information about an enterprise, in order to help them make better informed business decisions (Zeng et al 2009)."

Business Intelligence (BI) refers to various software solutions, including technologies such as ETL, Data warehouse, OLAP, Data mining &other reporting applications, share point server and web-enabled interface and methodologies needed to acquire the right information necessary for the business decision-making with the major purpose of enhancing the overall business performance on a marketplace. Data will be extracted using ETL (Extraction transformation and loading), stored in DW and generated reports (will be generated with the help of OLAP, DATA MINING and other reporting tools) can be accessed by the end-users through user interface. The BI tools were able to analyze the data for decision support in a fast and accurate manner.

**ETL Process:** ETL packages extract data from internal and external sources, eliminate data error and redundancies, and provide tailor data for access and analysis and load to DW. An important part of this process is data cleansing where variations in data schemas and data values from disparate transactional systems are resolved.

**DW (Data warehouse):** Collects relevant data into a repository, where it is organized and validated so it can serve decision-making objectives. The various stores of the business data are extracted, transformed and loaded from the transactional systems into the data warehouse.

**OLAP (online analytical process):** Depending on the organisations requirement one or more data cubes will be created. Each OLAP database contains specific number of cubes and dimensions. OLAP is a multidimensional model can then be created which supports flexible drill down and roll-up analyses (roll-up analyses create progressively higher-level subtotals, moving from right to left through the list of grouping columns. Finally, it creates a grand total).

**Data Mining:** Data mining tools for determining patterns, generalisations, regularities and rules in data resources.

**User Interface**: Easy and comfortable information access is mandatory for any BI .UI is a standard way to provide "single point of interaction" between users and BI solution.

According to Ranjan (2008): BI is the conscious, methodical transformation of data from any and all data sources into new forms to provide information that is business-driven and results-oriented. It will often encompass a mixture of tools, databases, and vendors in order to deliver an infrastructure that not only will deliver the initial solution, but also will incorporate the ability to change with the business and current marketplace (Ranjan 2008).Hocevar et al (2010), have found that the main categories of business intelligence benefits can be successfully linked to the defined long-term business strategy. The investment therefore helps the company achieve its strategic objectives which is one of the crucial criteria for deciding whether the investment in business intelligence is justified or not (Hocevar and Jaklic 2010).

BUSINESS INTELLIGENCE APPLICATIONS IN BANKING INDUSTRY

The way organizations approach a business intelligence project and the culture of the organization is as much of a factor of success as the integrated technology used to deliver the solution. BI would centralize the customer's information of bank, providing valuable insight (e.g. historical analysis, performance analysis, what if analysis, profit analysis, executive dashboards for managing customer relationship) throughout the organization, to improve the efficiency and provide better customer support. This would enable all the employees of the bank, to obtain all the relevant information from a single source, the BI systems, in order to carry out their business. Applications of BI in banks can be summarized as follows ((Bhasin 2006); (Katarina, et al ,2007), (Koh & Chan, 2002), (Decker, 1998))

**Banks performance (BP)**: Banks analyze their historical performance over time to be able to plan for the future. The key performance indicators include deposits, credit, profit, income, expenses; number of accounts, branches, employees etc.

**Marketing:** One of the most widely used areas of data mining for the banking industry is in marketing. The bank's marketing department can use data mining to analyze customer databases and develop statistically sound profiles of individual customer preferences for products and services. By offering only those products and services the customer really wants, the bank saves money on promotions and offerings that would otherwise be unprofitable.

**Risk Management (RM):** In a promptly changing and uncertain financial world, banking institutions need to rely more on fact-based actionable information, gleaned from ever-increasing data assets, to reduce risk wherever possible. BI is widely used for risk management in the banking industry. Bank executives need to estimate the reliability of their customers. Lack of knowledge regarding future customers may prove to be a great risk while offering new customers credit cards, extending existing customers lines of credit, and approving loans. Construct Credit scoring models to assess the credit risk of the loan applicants and construct fraud detection models to give signals of possible fraud transactions. Card theft analysis showed that number of transactions increases rapidly after the theft. By comparing expected average number or value of daily transactions, the authorization system can issue an early warning.

**Customer segmentation:** It is a method of grouping customers into classes that share similar patterns or characteristics. Organizations can use clustering to discover where similarities exist in their customer base so they can create and understand different groups to which they sell and market.

**Fraud Detection:** Being able to detect fraudulent actions is an increasing concern for many businesses; and with the help of data mining, more fraudulent actions are being detected and stopped. According to Decker, two different approaches have been developed by financial institutions to detect fraud patterns. In the first approach, a bank taps the data warehouse of a third party (potentially containing transaction information from many companies) and uses data mining programs to identify fraud patterns. The bank can then cross-reference those patterns with its own database for signs of internal trouble. In the second approach, fraud pattern identification is based strictly on the bank's own internal information (Decker, 1998).

**Customer Acquisition and Retention:** BI helps in acquiring and retaining customers in the banking industry. BI techniques can be used to study customers' past purchasing histories and know what kind of promotions and incentives to target customers. Also, if a branch has seen a number of people leave and go to competitors, BI can be used to study their past purchasing histories, and use this information to keep other customers from doing likewise. The findings can be used to prepare e-mail catalogues, target advertisements and promotion campaigns.

**Cross-selling:** The new mantra of marketing in banking is "the right product to the right customer at the right time". Construct models to predict the probability of purchasing certain products or services in order to facilitate cross-selling or up selling. Advantages of having correctly estimated probabilities are twofold: lowering the marketing campaign costs having a high response rate, and, more importantly, raising the quality of customer relations.

**Budget Planning:** Defining performance indicators from a specific area, calculated from the existing information from the system. We can follow the entire performance of the company or the performance of some groups as against others from the same area.

**Client Lifetime Value:** Customer lifetime value management estimates expected revenue from each customer in the future period. It is expected that a person with high education has higher income and is willing to meet the expense of additional products. The BI can build models for expected client lifetime value, so that bankers can treat clients accordingly, considering client's profitability based on his complete history.

The bank can start using BI for tracking business opportunities and developing customized solutions for its customers. Information extracted from BI needs to relate with business goals and objectives as well as business processes. Unfortunately, these aspects are poorly supported by BI. The developers of the BI have a data-centric viewpoint of business operations, rather than a process-centric perspective (Kubheka , 2007). KM System will play a role at this stage. KM system puts the information into a business context, improves the business decision-making and action-taking processing because the results become actionable. Flexibility and open architecture allow for easy expansion of the BI system. It is necessary in a situation when there are new informational needs or when an amount of information to be processed remarkably increases. Through KM system employees can send their feedback, need for improvement and add new information to the BI Solution. Implementation of BI applications usually takes time to develop and perfect, it also needs to modify the existing applications or add new applications as per the change in the market or business process. KM system helps to interact with the end-user during the development and enhancement phases to increase their satisfaction level with the system.

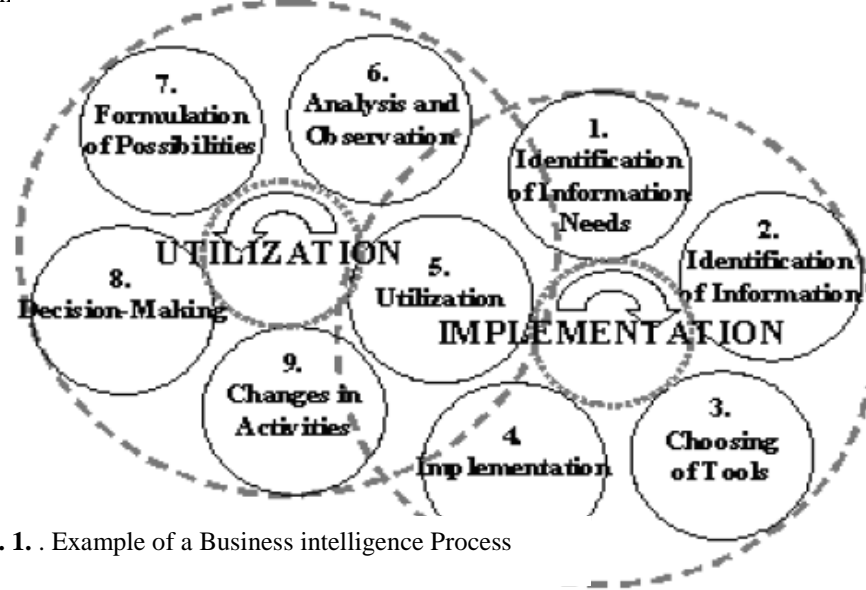

**Fig. 1.** . Example of a Business intelligence Process

As per Pirttimäki et al (2004), BI process concept is understood as a continuous and systematic method of action by which an organization gathers, analyses, and disseminates relevant business information to business activities and it should consist of two main processes: an implementation process and a utilization process (see Figure-1). Some phases of these processes can overlap each other. In the implementation process, business information needs are defined, relevant data and information are gathered, suitable tools are chosen, and finally the information gathered is stored. The utilization process has the following phases: utilization, analysis and observation, formulation of possibilities, dissemination of information and knowledge, decision-making, and changes in strategic and operational operations (Virpi, 2004). Knowledge sharing is intangible benefit for business intelligence.

**KNOWLEDGE MANAGEMENT**

End-users are required to use their experience/knowledge to make the decisions after receiving the information from BI. The decisions making process may involve the interaction with other users. KM enhances this interaction among the end-users by providing collaborative applications. In other words, KM system serves as a tool for collaborators to learn and exchange information regarding social and environmental actions taken within the organisation and by other organisations.

------------------Insert Figure. 2 somewhere here ------------------------

Nonaka and Takeuchi (1995) developed the knowledge spiral model to represent how tacit and explicit knowledge interact to create knowledge in an organization. The framework for a learning organization (see Figure 1) identifies four knowledge conversion processes or patterns: 1). Socialization (tacit to tacit), 2). Externalization (tacit to explicit), 3). Combination (explicit to explicit); and 4). Internalization (explicit to tacit)(Nonaka and Takeuchi, 1995).

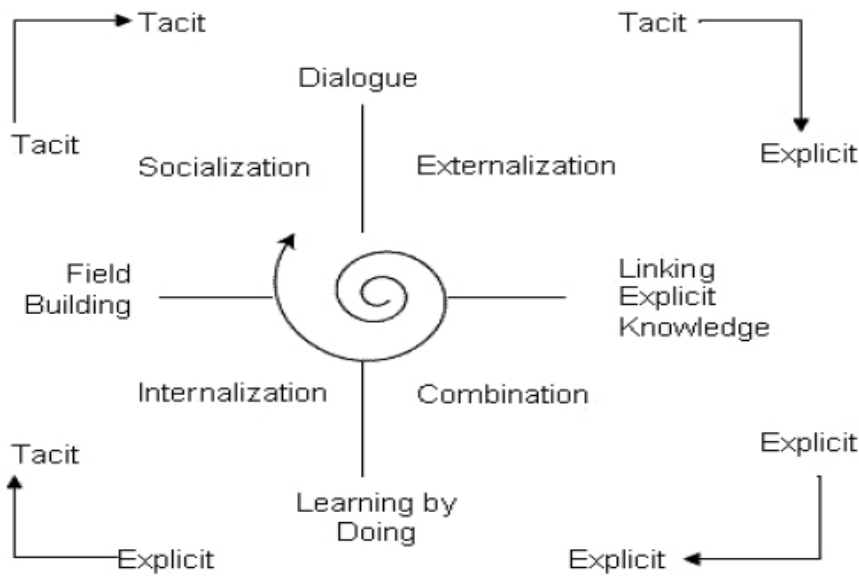

Fig.2. Nonaka KM Model

**Socialisation:** It is a way of capturing an individual's tacit knowledge. It occurs by sharing experiences and on-the-job learning. These experiences are shared openly and the individuals have to be motivated to be involved in the socialisation process. The knowledge conversion is from individual tacit to group tacit.

**Externalisation:** This is the process of converting group tacit knowledge into explicit knowledge. At this stage tacit knowledge is converted into models, prototypes and hypotheses. This knowledge can be used to come up with new products or processes.

**Combination:** This mode transforms explicit knowledge into further explicit knowledge by embedding knowledge into documents, manuals and so on.

**Internalisation:** This is the process of transforming explicit knowledge into tacit knowledge. Individuals internalise their experiences in performing specific functions or tasks. This mode is all about 'learning-by-doing'.

Knowledge management is the practice of adding actionable value to information by capturing tacit knowledge and converting it to explicit knowledge; by filtering, storing, retrieving and disseminating explicit knowledge; and by creating and testing new knowledge. In this context, tacit knowledge includes the beliefs, perspectives, and mental models so ingrained in a person's mind that they are taken for granted (Nonaka and Takeuchi, 1995). KM will have six dimensions: 1) creating of knowledge, 2) acquiring of knowledge, 3) organizing of knowledge, 4) saving of knowledge, 5) disseminating of knowledge and 6) applying of knowledge. Lee (2000) writes that: "inclusion of human's collaboration and help is the factor that distinguishes knowledge from corresponding data and information with it and this adds more value to the individual to whom knowledge is transferred". KM improves the knowledge activities performance, process performance, employee performance, market performance, and organizational performance (Cebi, Aydin and Gozlu, 2010). Data and information are the most important objects which computer processes and analyses. But now there are some people think of computer as knowledge processing machines. So we must think about their relation and difference.

- "Data" are the symbols, numbers, textual clauses, and other descriptive phrases or displays of measurements (e.g., evidence). They need to be processed into information through analysis and sort. Data is only the material of information.
- "Information" is built from the organization of data sets through quantitative and/or qualitative analysis that relate data sets, and can range from math equations, paragraphs, graphical illustrations, or images. Information also can be considered as an aggregation of data.
- "Knowledge" is created by applying experience to available measurements, data, and information. It is derived from information and context through judge.

## BANKS EXPERIENCE ON KNOWLEDGE MANAGEMENT

Maryam B et al (2010) have studied the KM practices and experiences of Iran banks. Their study shows that informal training is the main source of communication for sharing knowledge. Working on the other aspects such as IT systems, for the ease of strong and sharing experiences or lesson learned are useful. The study elaborates on capturing knowledge from industrial resources in three investigated banks, such as industrial associations, competitors, clients and suppliers. It showed that banks adopt themselves with the changing environment and can be more proactive than reactive (Maryam, Rosmini and Wan, 2010) . Today competitive benefits of strategic attempts along with knowledge management have relatively been recognized among all industries across the world. Research results give firm evidences and documents about this point that organizational culture has a positive relation with knowledge management and organizational benefits programs (Allame et al, 2010).D. CHATZOGLOU et al, Proposed an approach for integrating a system that utilizes decision support and KM to enhance the quality of the support provided to decision makers in a bank's loan department is presented. Some of the benefits of this new system include enhanced quality of support provided to bank managers in real time decision-making and KM functions. Furthermore, banking technologies contribute great benefits not only to banks themselves but also to their customers (e.g., convenience, security, improvements, better access to information, and an alternative to cash). Thus, new models which include all the factors mentioned in the above statements will provide more benefits and conveniences to banks and their customers (Chatzoglou, Vanezis and Christoforidis, 2005).The value of knowledge on bank's customers and products can erode over time. Since knowledge can get stale fast, the content in a knowledge management programme should be constantly updated, amended and deleted using results from regular survey of customers and Customer Satisfaction Index. Therefore, there is no endpoint to a knowledge management program. Like product development, knowledge management is a constantly evolving business practice which reflects the needs of banks' customers (Yarong & Ling 2006). True enterprise-wide KM solution cannot exist without a BI-based meta-data repository. In fact, a metadata repository is the backbone of a KM solution. That is, the BI meta-data repository implements a technical solution that gathers, retains, analyses, and disseminates corporate ''knowledge'' to generate a competitive advantage in the market, the intellectual capital (data, information and knowledge) is both technical and business-related (Marco, 2002). Techniques of knowledge discovery such as on line analytical processing and data mining, though they support the management of explicit knowledge, help in mastering the hidden knowledge of the individual in the decision making process. The decision making system can be observed as decision making based on rules and decision making based on skills and knowledge (Curko,Vuksic & Loncar,2009).

## INTEGRATION OF BI AND KM

Studies show that IT executives believe business managers do not understand what data they need and business managers reflect their belief that IT executives lack the business acumen to make meaningful data available. There is no easy solution to this problem; the beginning of the solution is for business managers and IT managers to pledge to work together on this question(Davenport and Harris, 2007).One of the factors for BI success is team work and business-IT alignment, and is necessary to support adoption and use of BI. Second involves implementation success outcomes when democratization or universal user adoption of BI has been achieved. In past studies measuring BI success, only a small portion of users had access to BI capabilities, successful outcomes can be realized while extending BI benefits to all users (Joseph, 2011). BI/MIS converts data into information, and then into knowledge that finally meets the needs of users of the system. KM system is on extracting knowledge from data and information. KM puts more emphasis on the knowledge itself and improves the utilization process of BI (Weidong, Weihui & Kunlong, 2010). To perform the Business Intelligence systems (BI) function effectively, such a system can be thought of as an intelligence cycle, representing a continuous process which is improved through feedback. This feedback process helps management to understand the end-user expectations, requirement and their usage pattern. The most critical element in the intelligence cycle is the determination of the intelligence requirements because this will determine the nature of the information system (Vasilopoulos, 2009).

Training end-users is important to improve knowledge and appropriate use of the system. General BI concepts, components, demonstration, and use are key training areas. Training should also include process changes and overall flow of information and integration (Bhatti, 2005).There is a positive relationship between collaborative culture, top management support and training and Business Intelligence success and this can be achieved through KMs. With the help of Business Intelligence and Knowledge Management, global businesses are increasingly successful. Knowledge management is deployed across a network of social and technical, human and material components. In the global economy, knowledge management is in fact a form of intercultural management (Albescu, Pugna & Paraschiv, 2009) and is designed to assist the interpretation of business cases by providing expertise, global domain knowledge.

Table 1. Difference between BI and KM

|  | BI | KM |
|---|---|---|
| **Sources** | Internal and external structured data sources. Data about suppliers, employees and customers etc. | Expert employees, Communities of interests / practices, organization, Market & Competitors structured/ unstructured data sources. |
| **IT** | Source systems, ETL ,DW,OLAP, Meta Data, Data Mining ,Statistical Analysis reporting and user interface | Document management, Web content management, enterprise knowledge portal , work flow, collaboration and e-learning |
| **Business Process** | Converts data into information & then into knowledge that finally meets needs of end-user. | Knowledge Sharing, Knowledge extraction, Knowledge communication, Knowledge application, and knowledge innovation. |
| **Deals with** | Explicit Knowledge, which is extracted from operational data. KPI, Process optimization, predict from internal and external data. | It deals with explicit as well as tacit knowledge. Informal, Formal, synergic and operational knowledge. |
| **Objective** | Identifies trends and patterns in structured data for developing new business strategies. Utilizes the massive data to discover the knowledge to provide competitive advantage. | Captures, stores, organizes, and distributes organizational knowledge and resources. It deals with the unstructured knowledge and tacit knowledge of the employees. |
| **Depends** | It depends on KM to receive feedback/experience from end-users and then to modify the solution, if required. | Depends on BI techniques to implement in an efficient way and explicit knowledge generated by BI. |

The benefits of integrating of BI with KM are 1) Ensure a real support in deploying successful business across the organization by smoothly managing multicultural teams of employees in providing highest quality products and global services to multicultural customers. 2) End-user preference and experience can be included in BI implementation, 3) provide better understanding on business context, interpretation results and training to the end-user. Even though both of them differ in their objectives and technologies used to develop them, together BI and KM can improve the organizational performance. Both BI and explicit knowledge deal with only a part of the NONAKA's- KM model .BI and KM integration assists today's managers for improved/ optimized decision making process by sharing data and information across the organization, getting the details from internal and external sources, forecasting the future trend and taking better decision.

As per Troy Hiltbrand (2010), adoption of BI with social practices can improve our ability to participate with information to produce significant and strategic corporate outcomes (Hiltbrand, 2010). BI focuses on explicit knowledge, but KM encompasses both tacit and explicit knowledge .Both concepts promote learning, decision making, and understanding. Yet, KM can influence the very nature of BI itself (Herschel & Jones 2005) .Hamid R. Nemati et al (2002), have proposed knowledge warehouse architecture as an extension to the BI model. The KW architecture will not only facilitate the capturing and coding of knowledge but will also enhance the retrieval and sharing of knowledge across the organization (Nemati et al, 2002) .As per Lingling Zhang et al (2009), there is no framework of knowledge management technology to well support analytical original knowledge generated from BI, which to some extent means that the way of incorporating knowledge derived from BI into knowledge management areas remains unexplored. They have proposed A4T transformation model involving data, rough knowledge, intelligent knowledge, and actionable knowledge, as well as the research direction, content and framework of future intelligent knowledge management (Zhang, Li and Shi, 2009) . There are three levels of integration between BI and KM (Baars 2005) : (1) Presentation level integration provides a horizontal integration with a joint user interface. (2) Data level integration provides the content of KM systems for BI processes by storing the related metadata into data warehouse. (3)System level integration provides distribution and re-utilization of BI analysis models by a knowledge management system.

## INTEGRATE BI AND KM IN BANKING INDUSTRY

R. I. Scott et al (2001) have shown the extensive role that domain knowledge plays in every step of the Business Intelligence Implementation process. In this case, the information was supplied by banking domain experts. The aim to integrate, ETL,OLAP, data mining, data warehousing, business modelling and high performance computing technologies to enable banks to increase profitability by the improved use of the vast amounts of customer-related data they hold. The interpretation of BI results is another step in the BI implementation process that relies heavily on domain knowledge. Often, this interpretation is based on the intuition of a domain expert and is therefore difficult to model (Scott et al, 2001). KM plays a role on all four processes of Nonaka KM model whereas BI plays a role on combinations, i.e., BI will transform the dirty data into explicit/codified knowledge.  But the experience of the end-users on BI usage can be called as tacit and KM will help end-users to share their experience by converting that into codified /explicit knowledge.BI deals with a subpart of the KM, without BI it is not possible to get that knowledge, so Integration of BI and KM will provide a more value.

BI system could be accessed by the employees, managers and executives of the bank in a hierarchical fashion, i.e., end-users from IT-Center and Central office can access the bank as a whole; users from different countries, zones/regions/branches can access the information which is related to their particular country, zone, region, or branch. Executives/employees from IT-center and Central Office can understand well about the technically terminology of BI, but not the users from regional, zonal or branch offices. Hence, users may not get the information they required and this process can take significant time depending on the change process initiated by the bank. However, banks can get positive result through 1) Proper guidance and continuous support from the IT and domain experts,2) Information sharing within the bank. For example Technical team members can provide solution for the technical problems and provide brief description about the critical reports on KM. Technical experts can visit the KM portal weekly to answer technical queries. Feed back of end-users can be received and modify existing BI system. The biggest problem for employees is to find the answer to specific questions, for this expert from each department of the bank can be asked,1) to share their innovative ideas about the information that is available within system and 2) to encourage users & promote the system. Knowledge and information sharing can be made the culture of the bank.

------------------ Figure 3 can be included some where here----------------------------------------

Integrated BI and KM provide the robust system with the capability of process-driven decision making. The processes are stored in process model base and their flexibility and reuse help enterprises improve the speed and effectiveness of business operations (Lee, 2000). Fig.3 helps us to understand the technical architecture to integrate BI and KM for improved decision making process and enhanced business performance. Employees may analyze potential growth and profitability of customer and reduce the risk exposure through more accurate financial credit scoring of their customers. This helps to solve problem related to churn management and helps to analyze why and how customers had left or are likely to leave. End-users can identify their most profitable customers and the underlying reasons for those customers' loyalty, as well as identify future promotional schemes for customers.

**ICICI BANK AS AN EXAMPLE**

In India, ICICI Bank's data warehousing capability is powered by Sybase IQ's unique column-based architecture, developed specifically for analytics and business intelligence.The foundation for ICICI Bank's wide-ranging CRM programs is a Sybase IQ-based data warehouse (The Bank initially used Teradata as its data warehouse platform and migrated to Sybase IQ a year ago.). The first iteration of the warehouse in 2000 generated a wealth of insights that enabled the bank to build customer intimacy, reduce churn, and offer cross-sell and up sell promotions. ICICI Bank deployed PowerCenter in 2003 as it embarked on the next phase of its warehouse, which would add data from five new sources, in addition to the initial three sources of retail banking, credit cards, and securities information. PowerCenter's easy-to-use drag-and-drop interface, native connectivity to a wide range of data sources, and standards-based architecture have helped ICICI Bank's internal IT personnel rapidly develop expertise in Informatica-based data integration (Chandana, 2008).

KM portal named 'Wiseguy' at ICICI India began on an experimental basis and carried on expanding and exploring, widening its ambit of operations. To develop 'WiseGuy', a team was put together encompassing KM, HR, technology and research with a brief to 'just do it'. Indeed they did and a beta version was ready within just three months. The requirement of relatively young age group of employees to achieve an understanding of the working culture and support provided by the top management had led to the progress of the concept of Knowledge Management. To a large extent, it is seen that the benefits of implementing 'Wiseguy' fulfils the needs of the technical and professional workforce by giving them a platform for airing their views, contributing as well as upgrading their expertise. It has also achieved in being a learning organization. KM at ICICI Bank was started in a non-dictatorial manner and its use is voluntary, but a programme of this nature cannot be expected to continue on momentum (Source: ICICI bank-Success story 2009).

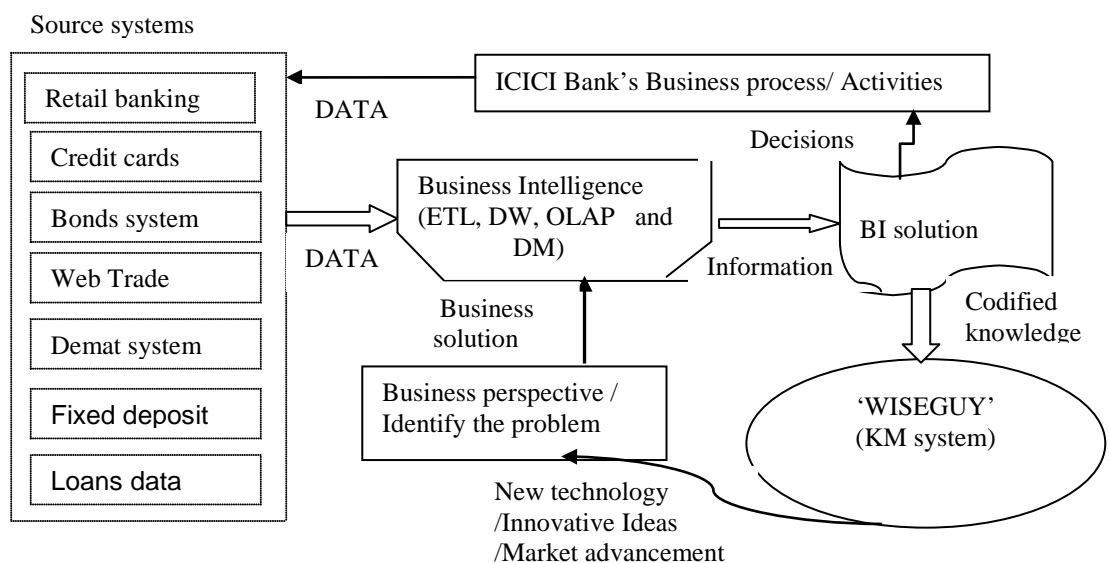

Figure-4

In order to be successful in today's dynamic business environment, ICICI has to continually improve and upgrade its BI system. Employees require information at all levels of the organization for ongoing decision making processes. Integration of BI and KM (Figure) increases the usage of the knowledge generated through BI system. This allows top management to understand the end-users perception and make further changes in BI system, if required. Though KM was started with the initiative taken by young age group employees with top management support, later the middle level employees also realized the benefits of KM and started using it. It may face problems in future as it is not getting upgraded in a strategic manner. They need to have a meta-data repository, which supports to maintain KM repository in a systematic way and helps users to find the required information in an effective manner.

**CONCLUSION**

Banks can use technology to improve their performance and they can get the sustainable competitive advantage. According to our study, we can conclude that proper integration of BI & KM can help bank to get wide benefits. It includes historical context, not just a shallow examination of what is apparent and easily accessible. Instead of nuggets or pockets of information from corporate databases, it provides a true 360º view of attitudes and behaviours, combines structured and unstructured data, meshes solicited and unsolicited feedback, and keeps a real-time pulse on business (Kadayam,2002). Banks will be able to manage explicit information and thereby transform the information to knowledge which in turn can help bank in making better decisions and lead them to be in a better position in contemporary business competitive environment. This integration will not only facilitate the capturing and coding of knowledge but also enhances the retrieval and sharing of knowledge across the bank to gain strategic advantage and sustain in competitive market.

**REFERENCES**

Albescu, F., Pugna, I., & Paraschiv, D. (2009). Cross-cultural Knowledge Management. *Informatica Economica*, 13(4), 39-50.

Allame, S., Nouri, B., Tavakoli, S., & Shokrani, S. R. (2011). Effect of Organizational Culture on Success of Knowledge Management System's Implementation (Case Study: Saderat Bank in Isfahan province). *Interdisciplinary Journal of Contemporary Research in Business*, 2(9), 321-346.

Baars H (2005). "Integration von Wissens management- und Business-Intelligence- Systemen –Potenziale". WM2005: Professional Knowledge Management – Experiences and Visions. Deutsches Forschungszentrum für Künstliche Intelligenz DFKI GmbH, pp 429-433

Bhasin, M.L. (2006), Data Mining: A Competitive Tool in the Banking and Retail Industries

Bhatti, T.R.(2005), "Critical Success Factors for the Implementation of Enterprise Resource Planning (ERP): Empirical Validation," *The Second International Conference on Innovation in Information Technology*).

Cebi, F., Aydin, O. & Gozlu, S. (2010). Benefits of Knowledge Management in Banking. *Journal of Transnational Management*, 15(4), 308-321. doi:10.1080/15475778.2010.525486

Chatzoglou, P. D., Vanezis, P., & Christoforidis, S. (2005). Knowledge Management Practices in the Financial Sector. *Knowledge-Based Economy: Management of Creation & Development*, 69-82.

Chandana G (2008), KNOWLEDGE MANAGEMENT IN INDIA : A CASE STUDY OF AN INDIAN BANK *The Journal of Nepalese Business Studies* Vol. V, No. 1, 2008, December Page: 37-49

Decker, P. (1998), Data Mining's Hidden Dangers. *Banking Strategies*, 6–14.

Hocevar, Borut; Jaklic, Jurij (2010).ASSESSING BENEFITS OF BUSINESS INTELLIGENCE SYSTEMS – A CASE STUDY. Management: Journal of Contemporary Management Issues,Jun2010, Vol. 15 Issue 1, p87-119.

ICICI bank-Success story (2009), ICICI Bank Improves IT Productivity and Systems Performance for Award- Winning Data Warehouse with Informatica Data Integration Platform. http://www.informatica.com /INFA_Resources/cs_icici_6806.pdf

Joseph, W. (2011). Business Intelligence Best Practices for Success. *Proceedings of the European Conference on Information Management & Evaluation*, 556-562.

Curko, K., Bach, M.P. , Radonic, G.; (2007),Business Intelligence and Business Process Management in Banking Operations. Proceedings of the *ITI 2007 29th Int. Conf. on Information Technology Interfaces,* June 25-28, 2007, Cavtat, Croatia

Koh Hian, C., & Chan Kin Leong, G. (2002). Data Mining and Customer Relationship Marketing in the Banking Industry. *Singapore Management Review*, 24(2), 1

K. Curko, V. Bosilj Vuksic, A. Loncar(2009),The Role of Business Process Management Systems and Business Intelligence Systems in Knowledge Management. INTERNATIONAL JOURNAL OF COMPUTERS AND COMMUNICATIONS Issue 2, Volume 3, 2009

Kadayam, S. (2002). New Business Intelligence: The promise of Knowledge Management, the ROI of Business Intelligence.

Kubheka, NSP(2007),How to leverage information to improve business performance in a financial services company-Research Report

Lee, James Sr. (2000). Knowledge Management: The Intellectual Revolution. *IIE Solutions, 32*, 34-37

Lingling Zhang,Jun Li ,Yong Shi (2009),Foundations of intelligent knowledge management. Human Systems Management, 2009, Vol. 28 Issue 4, p145-161.

Marco, D. (2002). The key to knowledge management. http://www.adtmag.com/ article.asp?id= 6525

Maryam B, Rosmini O and Wan K (2010). Knowledge Management and Organizational Innovativeness in Iranian Banking Industry.Proceedings of the International Conference on Intellectual Capital, Knowledge Management & Organizational Learning; 2010, p47-60, 14p

Nonaka, I. (2007). The Knowledge-Creating Company. *Harvard Business Review*, 85(7/8), 162-171

Nemati, H. R., Steiger, D. M., Iyer, L. S., & Herschel, R. T. (2002). Knowledge warehouse: an architectural integration of knowledge management, decision support, artificial intelligence and data warehousing. *Decision Support Systems*, 33(2), 143

Petrini, M. & Pozzebon, M. (2009). "Managing sustainability with the support of business intelligence: Integrating socio-environmental indicators and organizational context." The Journal of Strategic Information Systems. 18(4), 178-191.

Ranjan, J. (2008): Business justification with business intelligence, The Journal of Information and Knowledge Management Systems, Vol. 38, No.4, pp. 461-475

Richard T.Herschel, Nory E.Jones (2005) ," Knowledge management and business    intelligence: the importance of integration", Journal of Knowledge Management, Vol. 9 No. 4, pp. 45-55

R. I. Scott S. Svinterikou C. Tjortjis J. A. Keane (2001). Experiences of using Data Mining in a Banking Application

Thomas H. Davenport andJeanne G. Harris, Boston (2007), Competing on analytics: The new science of winning.. Harvard Business School Press, 2007, 240 pp., ISBN-13: 978-1-4221-0332-6

Troy Hiltbrand (2010). Social Intelligence: The Next Generation of Business    Intelligence ,Business Intelligence Journal, 2010, Vol. 15 Issue 3, p7-13, 7p

Virpi Pirttimäki (2004). The Roles of Internal and External Information in Business Intelligence. FRONTIERS OF E-BUSINESS RESEARCH

Vasilopoulos(2009). Development of a Competitive Business Intelligence System Proceedings of Northeast Business & Economics Association (NBEA)

Yarong, C., & Ling, L. (2006). Deriving information from CRM for knowledge management—a note on a commercial bank. *Systems Research & Behavioral Science*, 23(2), 141-146. doi:10.1002/sres.756
Zeng, L., Xu, L., Shi, Z., Wang, M. and Wu, W. (2007), 'Techniques, process, and enterprise solutions of business intelligence', 2006 IEEE Conference on Systems, Man, and Cybernetics October 8-11, 2006, Taipei, Taiwan.

Zhao Weidong; Dai Weihui; Yang Kunlong (2010) , "The relationship of business intelligence and knowledge management," Information Management and Engineering (ICIME), 2010 The 2nd IEEE International Conference on , vol., no., pp.26-29, 16-18 April 2010

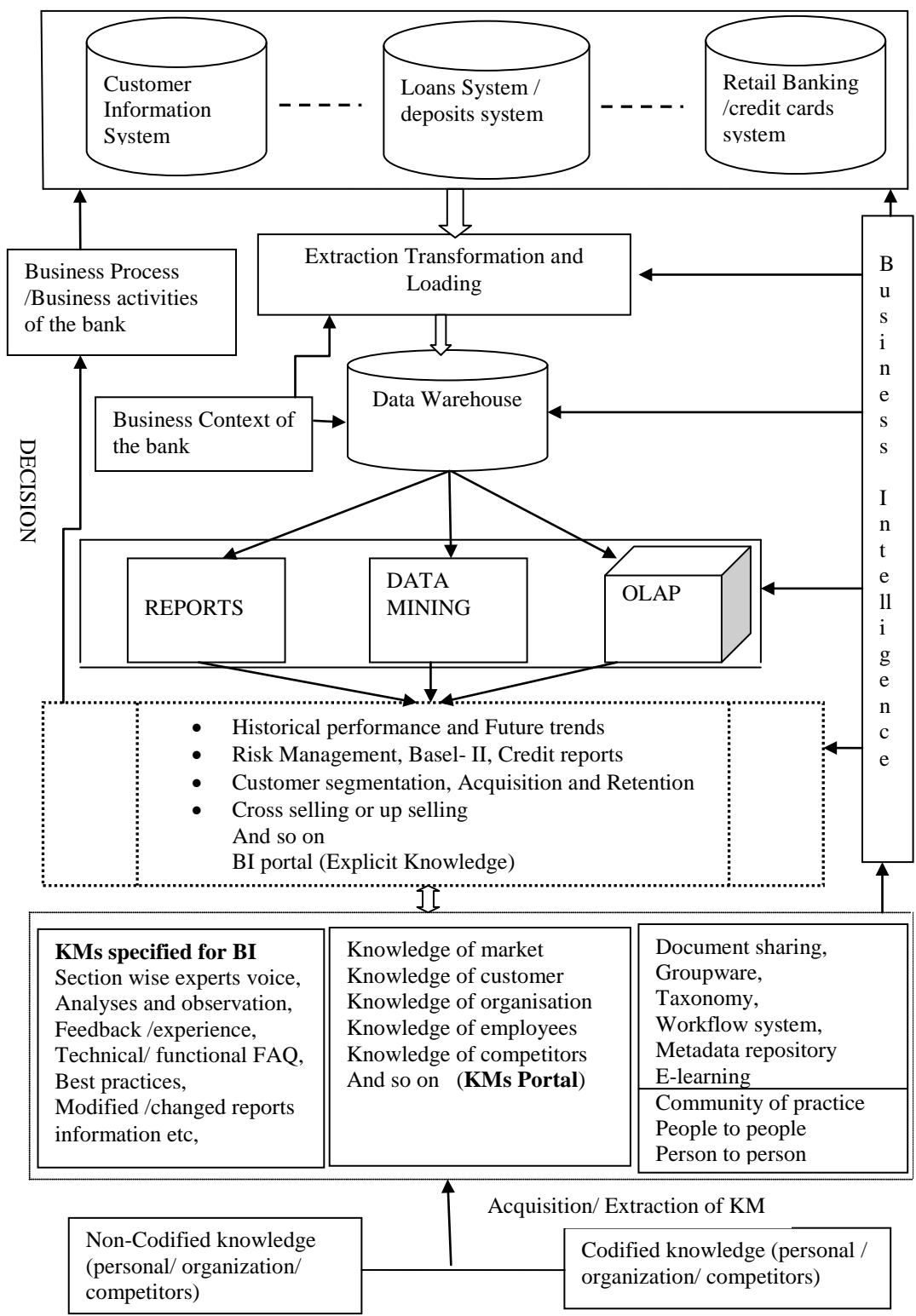

Figure.3   Technical Architecture to Integrate BI and KM in banks

**G. Koteswara Rao** received B.Sc degree in Mathematics, statistics and computer science and M.Sc in Mathematics from IIT KANPUR. He had worked with HCL as a Research Engineer from Oct-2007 to Oct- 2009. Currently he is working in IIM Indore as a Research Associate in Information Systems and Information Technology area. He has published papers in the areas like, e-governance, e-democracy, Banking Sector, Business Intelligence, Knowledge Management and Text Mining. He has also reviewed papers for ACITY-2011, CCSEIT-2011 and AOM Annual Meeting-2011.

**Roshan Kumar** holds Master of International Business from La Trobe University, Melbourne and a Bachelor of Engineering in Polymer Engineering from Birla Institute of Technology, India. He was an Assistant Systems Engineer at Tata Consultancy Services, India. He is presently working as a Teaching Associate at the Department of Communications, Indian Institute of Management, Indore , India.